\documentclass[letterpaper,11pt]{article}
\pdfoutput=1 
\usepackage{fullpage}
\usepackage{amsmath}
\usepackage{amssymb}
\usepackage{graphicx}
\usepackage{mathrsfs}
\usepackage{wrapfig}
\usepackage{boxedminipage}
\usepackage{epsfig}
\usepackage{setspace}
\usepackage{subcaption}
\usepackage{float}
\usepackage{hyperref}
\usepackage{enumerate}
\usepackage{cite}
\usepackage[font=footnotesize,labelfont=rm]{caption}

\newcommand{\reef}[1]{(\ref{#1})}
\newcommand{\beq}{\begin{equation}}
\newcommand{\eeq}{\end{equation}}
\newcommand{\bea}{\begin{eqnarray}}
\newcommand{\eea}{\end{eqnarray}}

\begin{document}
\begin{flushright}
{\tt arXiv:1709
}
\end{flushright}

\bigskip
\bigskip
\bigskip
\bigskip

\bigskip
\bigskip
\bigskip
\bigskip

\begin{center}
{\Large
{\bf Benchmarking Black Hole Heat Engines, II
}
}
\end{center}

\bigskip
\begin{center}
{\bf  Avik Chakraborty and
Clifford V. Johnson}

\end{center}

\bigskip
\bigskip

\begin{center}
  \centerline{\it Department of Physics and Astronomy }
\centerline{\it University of
Southern California}
\centerline{\it Los Angeles, CA 90089-0484, U.S.A.}
\end{center}

\bigskip
\centerline{\small \tt avikchak, johnson1,  [at] usc.edu}

\bigskip
\bigskip
\bigskip

\begin{abstract} 
\noindent  
We extend to non--static black holes our benchmarking scheme that allows for  cross--comparison of the efficiencies of asymptotically AdS black holes used as working substances in heat engines.  We use a circular cycle  in the $p{-}V$ plane as the benchmark cycle. We study Kerr black holes in four spacetime dimensions as an example.  As in the static case, we find an exact formula for the benchmark efficiency in an ideal--gas--like limit, which  may serve as an upper bound for  rotating black hole heat engines in a thermodynamic ensemble with fixed angular velocity. We use the benchmarking scheme to compare Kerr to static black holes charged under Maxwell and Born--Infeld sectors. 
\end{abstract}
\newpage \baselineskip=18pt \setcounter{footnote}{0}

\section{Introduction}
\label{Intro} 

\begin{figure}[h]
\centering
\begin{subfigure}[b]{0.25\textwidth}
\includegraphics[width=1.6in]{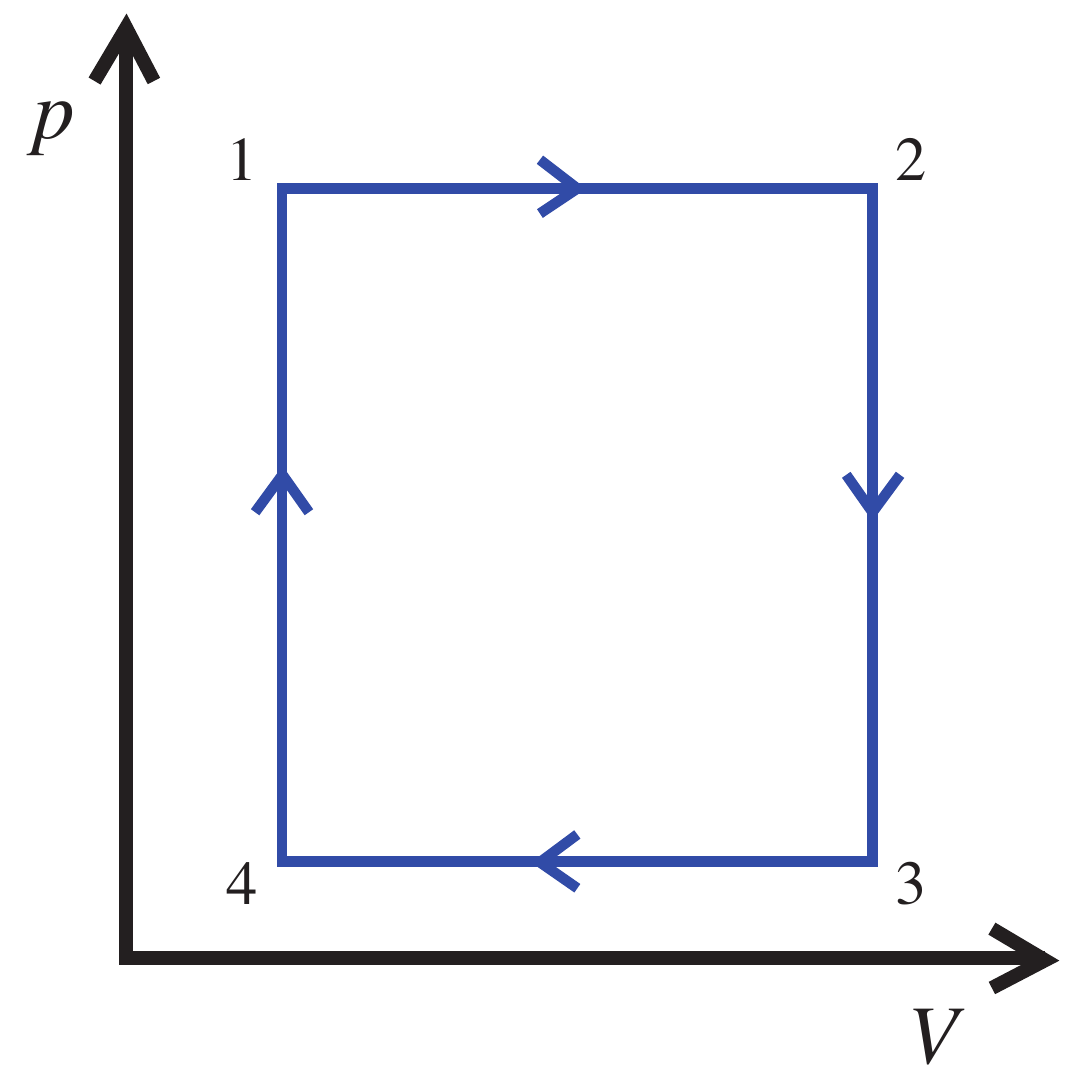} 
   \caption{\footnotesize  }   
 \label{fig:WE}
\end{subfigure}
\qquad\qquad
\begin{subfigure}[b]{0.25\textwidth}
\includegraphics[width=1.6in]{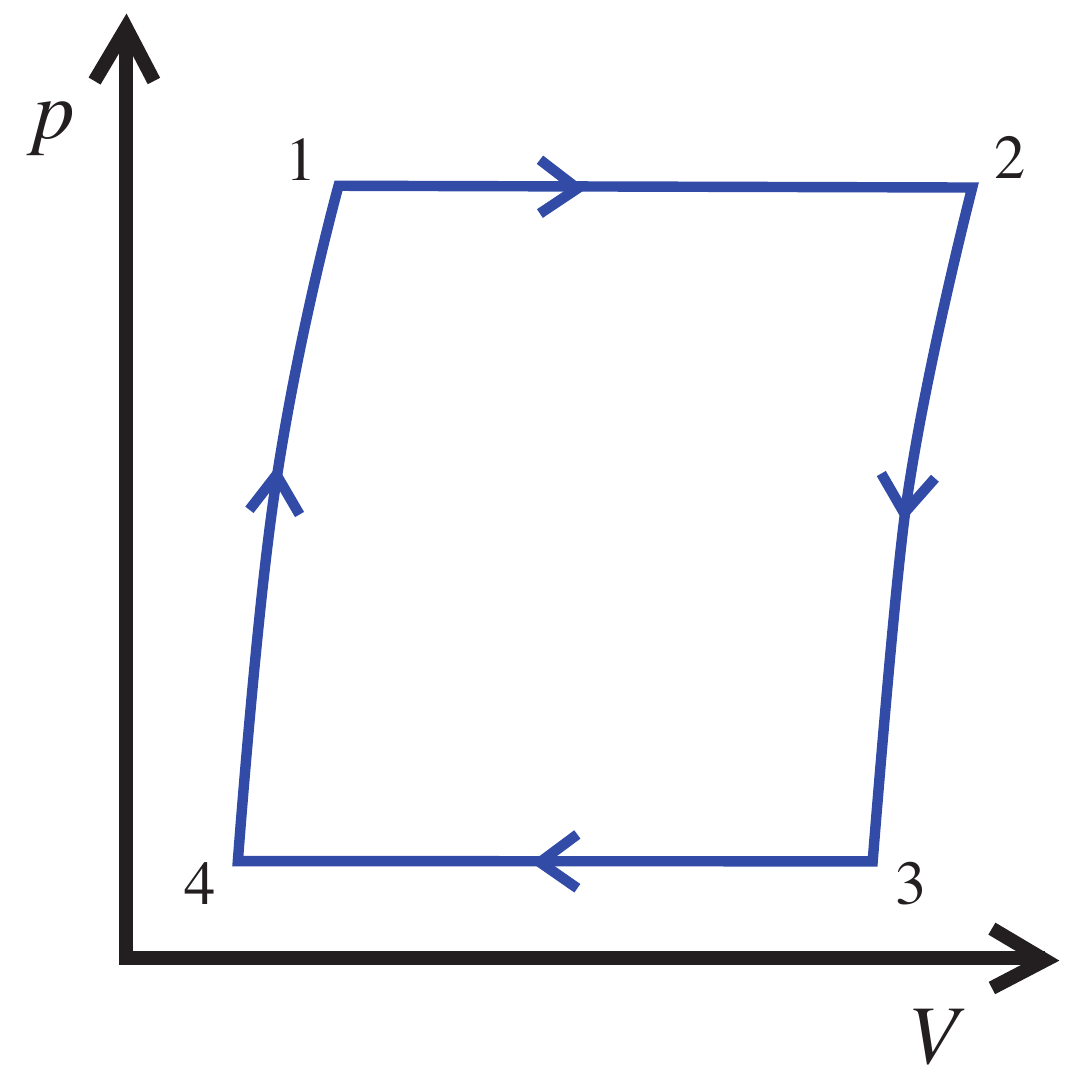} 
   \caption{\footnotesize  }   
\label{fig:NC}
\end{subfigure}
   \caption{\footnotesize (a) Prototype engine for static black hole. (b) Prototype engine for rotating black hole, where steps 2$\to$3 and 4$\to$1 are chosen to be adiabats.}   
   \label{fig:cycles}
\end{figure}

The classic black hole thermodynamics framework \cite{Bekenstein:1973ur,Bekenstein:1974ax,Hawking:1974sw,Hawking:1976de} can be extended\footnote{See {\it e.g.,} refs. \cite{Wang:2006eb,Sekiwa:2006qj,LarranagaRubio:2007ut,Kastor:2009wy},  the early work in refs. \cite{Henneaux:1984ji,Teitelboim:1985dp,Henneaux:1989zc}, and the reviews in ref. \cite{Dolan:2012jh,Kubiznak:2016qmn}.} by allowing the cosmological constant to be dynamical. It supplies a pressure {\it via} $p=-\Lambda/8\pi$, along with its conjugate volume\footnote{Here we are using geometrical units where $G,c,\hbar, k_{\rm B}$ have been set to unity.} $V$. In this formalism,  where   mechanical work  is possible because of  the $pdV$ term now appearing in the First Law of thermodynamics,  heat engines  are a natural concept \cite{Johnson:2014yja}, especially when working with negative cosmological constant\footnote{The mechanical work $W=\int pdV$ here  can be interpreted \cite{Johnson:2017hxu} as a change of the  overall energy of a spacetime since the volume ({\it i.e.,} the presence of the black hole itself) removes removes a portion of it from the standard energy integral. Recall that $p$ sets an energy density {\it via} $\Lambda$'s equation of state $\rho=-p$ and so a positive change $dV$ results in an energy gain $|\rho| dV$ \cite{Kastor:2009wy}. The heat flows $Q_H$ and $Q_C$ into and out of the engine can be considered as from and to non--backreacting heat baths of radiation  filling the spacetime, as is traditional in black hole thermodynamics. See {\it e.g.} ref. \cite{Hawking:1982dh}.  These devices were named  ``holographic heat engines'', since for  negative cosmological constant ({\it i.e.} with positive pressure), an engine cycle  represents a journey through a family of holographically dual\cite{Maldacena:1997re,Witten:1998qj,Gubser:1998bc,Witten:1998zw,Aharony:1999ti} non--gravitational  field theories (for a large number of colours, $N_c$) defined in one  dimension fewer. The holographic aspects will not be the focus of this paper.}. The equation of state of the working substance  is supplied by  the black hole in question in the form of the relation between its temperature~$T$, the background cosmological constant, and its other parameters such as horizon radius~$r_+$. The engine  is defined using a  closed cycle in state space during which there is a net input heat  flow $Q_{H}$, a net output heat flow $Q_{C}$, and a net output work $W$ such that $Q_{H}=W+Q_{C}$. The efficiency of the cycle, $\eta=W/Q_{H}=1-Q_{C}/Q_{H}$, is determined by the equation of state of the system and the choice of cycle in the state space.  The cycle given in Figure~\ref{fig:cycles}(a) is a natural choice for static black holes. This is because, in that case entropy and volume, both purely powers of $r_{+}$, are not independent, and hence the isochores are adiabats, ({\it i.e.,} $C_V=0$ \cite{Dolan:2012jh}), making all the heat flows take place on the isobars, simplifying the analysis\cite{Johnson:2015ekr,Johnson:2015fva} of the efficiency\footnote{Refs. \cite{Belhaj:2015hha,Sadeghi:2015ksa,Caceres:2015vsa,Setare:2015yra,Zhang:2016wek,Bhamidipati:2016gel,Wei:2016hkm,Sadeghi:2016xal,Wei:2017vqs,Hendi:2017bys,Mo:2017nes,Bhamidipati:2017nau,Xu:2017ahm,Liu:2017baz,Mo:2017nhw,Johnson:2017hxu,Johnson:2017ood} have since done further studies of such heat engines in various contexts.}.

In general, the calculation of efficiency is a difficult task to perform exactly and one adopts various approximation schemes (such as  high temperature limits) to proceed, but ref. \cite{Johnson:2016pfa} showed that for this cycle, a simple exact formula can be derived for it:
\begin{equation}
\label{eq:exact}
\eta=1-\frac{M_3-M_4}{M_2-M_1}\ ,
\end{equation}
where $M_i$ is the mass of the black hole (which is also the enthalpy\cite{Kastor:2009wy})  evaluated at the $i$th corner of the cycle. Moreover, since two such cycles can be placed together (sharing a common edge) to give a larger  cycle, the simple formula can be used in an algorithm for computing the efficiency of an engine defined by an arbitrary cycle shape in the $(V,p)$ plane, by tessellating the plane using  the basic cycle as a unit cell \cite{Johnson:2016pfa}. Only the  edges of the cells that intersect the cycle path will contribute to the efficiency, and  since the heat flows are on the isobars only, there are only two types of cell that contribute: ``hot'' cells that are on upward--facing parts of the cycle (therefore contributing to~$Q_H$) and ``cold" cells otherwise (contributing to $Q_C$), and hence:
\begin{equation}
\eta=1-\frac{\sum_j(M_{3}^{(j)}-M_{4}^{(j)})}{\sum_k(M_{2}^{(k)}-M_{1}^{(k)})}\ ,
\label{eq:algorithm}
\end{equation}
where the additional labels $j,k$  are for cold and hot cells respectively. As the number of cells used in this prescription increases (or equivalently, the cell size decreases), the accuracy of the algorithm increases.

In ref. \cite{Chakraborty:2016ssb}, this algorithm was used in a  prototype benchmarking scheme for comparing efficiencies of different kinds of heat engines. The idea was that since certain special shapes of cycle might be more advantageous  for  certain kinds of black hole ({\it e.g.} having isochors favours static black holes since $C_V=0$), it makes sense to use a benchmark shape that is equally disadvantageous to all equations of state. A circle was the natural choice\footnote{It's a circle in our choice of units for $p$ and $V$, which is part of the choice of benchmarking scheme if making comparisons.}: ($(p-p_c)^2+(V-V_c)^2=L^2$, {\it i.e.,} centered at $(p_c,V_c)$ and of radius $L$) and the algorithm gave a means by which $\eta$ for each case could be computed and compared. 

Proceeding in this way produced another useful outcome: In an ``ideal gas" limit, where black hole equations of state take the form $p V^{1/(D-1)}\sim T$, the mass becomes $M=pV$, and the entire scheme becomes a simple exactly solvable geometry problem with the result:
\begin{equation}
\label{eq:cjbound}
\eta=\frac{2\pi}{\pi+4p_c/L}\ ,
\end{equation}
where  $L$ is the circle's radius and $p_c$ $(\le L)$ is the pressure at the centre of the circle. This suggests a universal (dimension independent) bound on the efficiency in this benchmarking scheme, coming from the ideal gas sector. (It is stronger than the bound presented by the Second Law of thermodynamics through the Carnot efficiency.)

It is natural to wonder how all of this generalizes to cases for which $C_V\neq0$, {\it i.e.,} if the constant volume lines are not adiabatic curves. Rotating black holes fall into this category, as non--zero rotation parameter $a$ enters the formula for the thermodynamic volume and entropy in such a way as to make them independent from each other. This was the point of the current project, and in this paper we present the details of an interesting generalization of our previous results. While this project was near completion, a   generalization that is different from the approach we took appeared in the literature \cite{Hennigar:2017apu}. That paper also looks at the $C_V\neq0$ case, but instead opts to keep the prototype cycle in figure~\ref{fig:cycles}(a) made of isobars and isochores. A generalization of the exact efficiency formula \reef{eq:exact} can be readily written down, and they study  the Kerr (and other) black holes using it. In the benchmarking circle, the paper revisits the $C_V=0$ case and derives an analytic expression for a {\it lower} bound  on $\eta$. While not as universal (and dimension independent) as our upper bound~\reef{eq:cjbound}, it may be useful  in further studies of benchmarking. Their approach using the rectangular cycle for $C_V\neq0$ does not extend to a simple algorithm for solving arbitrary shapes, however, and so they study $C_V\neq0$ benchmarking by direct numerical evaluation of the heat around the circle.

Since our interest was primarily in generalizations of the tessellation algorithm  for the purposes of computing the efficiency of the benchmarking circle, we took a different approach, using two isobars and two adiabats to form our basic engine, giving a new prototype cycle that looks like that shown in figure~\ref{fig:cycles}(b). With this choice,  the heat flows are again just along the top and bottom paths of the cycle, and the simple algorithm that involves tessellating the $(V,p)$ plane with them in order to determine the efficiency for an arbitrary shape can be implemented again. The dependence of $S$ on the various parameters can be used, in combination with the equation of state, to determine the equations for the adiabatic curves at each step in the algorithm in such a way as to avoid needing to solve for the adiabatic curves in closed form (which is in general hard to do). We describe this generalized scheme in section \ref{sec:benchmarkgeneral} and then apply it to the example of the Kerr black hole in section~\ref{sec:kerr}. 
  
  As with our previous work \cite{Chakraborty:2016ssb}, it is possible to ignore the tessellation algorithm and do a direct numerical integration of the heat in order to compute the efficiency of the circle cycle. However, proceeding in that way would have meant missing the opportunity to derive the exact formula~(\ref{eq:cjbound}). Indeed, pursuing the generalization of the tessellation method for the rotating case, we found a generalization of that exact formula  by again taking an ideal gas like limit. This formula again suggests itself as an upper bound on  rotating  cases (in the fixed angular momentum ensemble that we'll be working in; see section~\ref{sec:details}). This is presented in section~\ref{sec:exact}.

\section{The Generalized Benchmarking Cycle}
\label{sec:benchmarkgeneral}

To generate the tessellation of the benchmarking cycle with unit cells of the form shown in figure~\ref{fig:cycles}(b), the procedures of ref.\cite{Chakraborty:2016ssb} need some modification. The benchmarking path is still a circle with center at $(V_c\,,p_c)$ and radius $L$, of course, but the core tessellating grid will now be $2N\times N$, to accommodate the bending of the adiabats. Here $N$ is again chosen to be even, since it gives equal numbers of cells in the upper and lower halves of the circle. Having chosen the origin and the radius (denoted $L$) of the circle in the $(V,p)$ plane, the lattice is constructed as follows. We started from the top left corner of the grid. The vertical decrement for pressure will be $\Delta p=2L/N$, and we have $N$ rows as before, resulting in an exact fit of the circle between the top most and bottom most isobars, along the vertical direction. In the horizontal direction, pressure is constant, and we have generated $2N$ columns in stead of $N$, {\it i.e.}, we have extended the length of the grid along this direction. A similar increment is used for $\Delta V$ to locate the corners of the tops of the cells in the top row. Since we have $2N$ columns now, the total length of the top isobar is $4L$ as opposed to our static case where it was $2L$. Next, for the adiabatic segments, setting $dS=0$ gives an equation that can be used to determine the required $\Delta V=F(V,p)\Delta p$ where $F(V,p)$ may also depend upon other parameters (such as $a$ or $J$ in the rotating black hole case). This determines all the steps needed to move vertically and horizontally, and the grid can be readily generated. For black hole solutions, it is often most useful to write quantities in terms of the horizon radius $r_{+}$, so we computed $r_{+}$ at each corner of the tops of the top cells in the top row using the explicit equation for $V$ in terms of $r_{+}$ and $p$ since its $(V,p)$ coordinates are known, with other parameters being held fixed. To generate the corners of the bottoms of the cells in the top row, we can now use $\Delta V$ because $\Delta V$ is a function of $V$, $p$ and $r_{+}$ with all the variables known. Corners of the tops of the cells in the next row are same as those of the corners of the bottoms of the cells in the top row. So we can start from those corners and compute $\Delta V$ using the same trick as before which gives us the corners of the bottoms of the cells in the second row and this process continues until we reach the $N$-th row. At the end of the process, $r_{+}$ at each corner is known (along with $p$ and $V$), and can be readily used to evaluate mass $M$, temperature $T$ and other quantities at each corner. Identifying hot and cold cells is done using the same method as in \cite{Chakraborty:2016ssb} and the efficiency $\eta$ (for that value of $N$) is readily computed. As described above, the circle fits exactly on the core $N\times N$ tessellating grid only when the adiabats are vertical (the static black hole case). Since the adiabats bend away from the vertical, the core grid was extended to $2N\times N$ in order to ensure that the circle fits on it.  

As $N$ increases, $\Delta p$ and $\Delta V$ get smaller, resulting in more accurate fits to the exact adiabats, and also the hot cells and cold cells make a better fit to the circle, so  $\eta$ becomes more accurate for large $N$. Since the algorithm computes $T$ at each  corner  of  all the hot cells and cold cells, it is easy to determine the maximum and minimum temperatures on  the benchmarking cycle in order to compute the Carnot efficiency ($\eta_{C}$) of the engine as a  check our results for $\eta$.

\begin{figure}[h]
\begin{center}
{\centering
\includegraphics[height=3.5in]{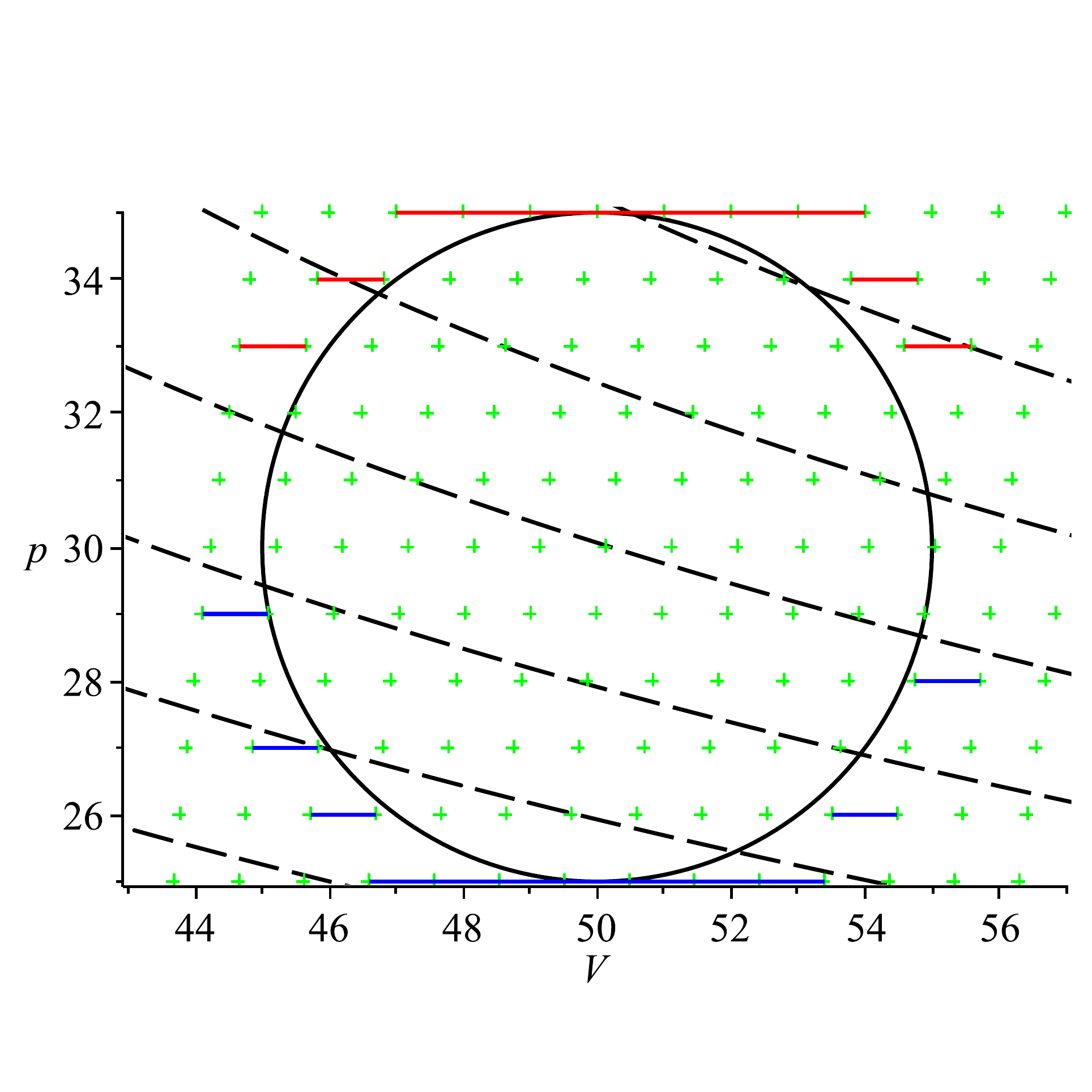} 
   \caption{\footnotesize  A sample benchmarking circular cycle for $N=10$, showing the tessellation along isobars and adiabats. The equation of state comes from $D=4$ Kerr--AdS black holes. Red lines represent hot cells and blue lines represent cold cells. As $N$ increases, these lines converge to the boundary of the circle. Black dashed lines show a family of isotherms. We worked at fixed $a$ here, with $a=0.04$.}   \label{fig:CCN10}
}
   \end{center}
\end{figure}

Figure~\ref{fig:CCN10} shows an example for the $D=4$ Kerr black hole for $N=10$ where  we worked at fixed rotation parameter $a=0.04$. (See the next section.) The benchmarking circular cycle is shown, with $(V_c=50, p_c=30)$ as the origin and $L=5$ as the radius. The  crosses (green) show the points of the  tessellation. The black dashed lines show a few sample isotherms.

\section{Benchmarking Kerr--AdS}
\label{sec:kerr}

\subsection{Background Thermodynamics}

We will work in $D=4$ dimensions in this paper. Though our approach should ideally work in any~$D$, for $D\geq5$ the relevant thermodynamic quantities become more complicated and the process proved to be beyond the numerical capabilities of our computers. (There will be an exact result in section~\ref{sec:exact} that will be $D$--independent, however.) Our purpose for this paper is to generalize/modify the previous algorithm for static cases to incorporate rotation. We will then compare the result with  non--rotating cases.

We use the $D=4$ Einstein--Hilbert action:
\bea \label{EHaction}
I=\frac{1}{16\pi}\int d^4 x\sqrt{-g}\Big(R-2\Lambda \Big) \ ,
\eea where the cosmological constant $\Lambda=-3/l^2$ sets a length scale $l$. The Kerr--AdS solution  in Boyer-Lindquist coordinates is given by:
\bea \label{metric}
ds^2=-\frac{\Delta_{r}}{\rho ^2}\Big[dt-\frac{a \sin^2 \theta}{\Xi} \ d\phi \Big]^2+\frac{\rho ^2}{\Delta_{r}} \ dr^2+\frac{\rho ^2}{\Delta_{\theta}} \ d\theta ^2+\frac{\Delta_{\theta} \ \sin^{2}\theta}{\rho ^2} \Big[a dt-\frac{r^2+a^2}{\Xi} \ d\phi \Big]^2 \ , 
\eea 
where,
\bea
\rho^2 =r^2+a^2 \ \cos^2 \theta \ ,\quad  \ \Xi =1-\frac{a^2}{l^2} \ ,
\eea
\bea \label{radial}
\Delta_{r}=(r^2+a^2) \ \Big(1+\frac{r^2}{l^2}\Big)-2mr \ ,\quad \ \Delta_{\theta}=1-\frac{a^2}{l^2}  \cos^2 \theta \ .
\eea
Here $a$ and $m$ are rotation and mass parameters, respectively. Note that this solution is valid only for $a<l$ and becomes divergent in the limit $a \rightarrow l$. The event horizon is at  the largest real root,~$r_+$, of  $\Delta_{r}=0$. The horizon area $A$ and entropy $S$ are  given by:
\bea \label{entropy}
A=\frac{4 \pi (r^2_{+}+a^2)}{\Xi} \ , \quad \ S=\frac{A}{4}\ .
\eea
Analytically continuing the metric by $t \rightarrow i \tau $ and $a \rightarrow ia$ gives the Euclidean section and if  we  identify $\tau \sim \tau + \beta$ and $\phi \sim \phi + i\beta \Omega_{H}$, where $\Omega_{H}$ is the angular velocity of the event horizon
\bea
\Omega_{H}=\frac{a \ \Xi}{r^2_{+}+a^2} \ .
\eea
and $\beta$ is the inverse  temperature:
\bea \label{temp}
\beta^{-1}=T=\frac{1}{2 \pi r_{+}} \left[\frac{(a^2+3r^2_{+})\Big(\frac{r^2_{+}}{l^2}+1\Big)}{2(a^2+r^2_{+})}-1 \right] \ ,
\eea
then the geometry is free of conical singularities. Note that the temperature vanishes at the following extremal value of $a$: 
\bea \label{eq:extremal}
a^2_{\rm ext}=\frac{r^2_{+}\Big(1+\frac{3r^2_{+}}{l^2}\Big)}{1-\frac{r^2_{+}}{l^2}}\ . 
\eea
The angular velocity measured at spatial infinity is what enters the First Law \cite{Cvetic:2010jb} (see below) and is:
\bea \label{AngVelSp}
\Omega=\Omega_{H}+\frac{a}{l^2}=\frac{a\Big(1+\frac{r^2_{+}}{l^2}\Big)}{r^2_{+}+a^2} \ .
\eea
The mass, angular momentum and thermodynamic volume  are \cite{Cvetic:2010jb}:
\bea \label{volume}
M=\frac{m}{\Xi^2} =
\frac{1}{2r_{+}\Xi^2}(r^2_{+}+a^2)\Big(1+\frac{r^2_{+}}{l^2}\Big)\ , \ J=\frac{ma}{\Xi^2} \ , \quad {\rm and} \quad \ V=\frac{r_{+}A}{3}\left[1+\frac{a^2\Big(1+\frac{r^2_{+}}{l^2}\Big)}{2 r^2_{+} \Xi}\right]\ .
\eea

The First Law of black hole thermodynamics for a rotating black hole is \cite{Cvetic:2010jb}:
\bea \label{FirstLaw}
dH=TdS+Vdp+\Omega dJ \ ,
\eea
where $\Omega$ is the angular velocity and $J$ is the angular momentum. The enthalpy $H$ is simply the mass of the black hole, usually written as a function of $r_{+}$ and $p$, parametrized by $a$, where, the rotation parameter $a=J/M$ (in $D=4$).

\subsection{Key Elements of the Tessellation}
\label{sec:details}
It is possible to explicitly rewrite all the thermodynamic quantities in terms of the angular momentum $J$, and  it  is indeed a natural thermodynamic variable to hold fixed. On the other hand for the purposes of exploring the generalized tessellation we (for simplicity) choose to work with  fixed $a$, amounting to  fixing $J$'s conjugate, the angular velocity at infinity. (Note that the benchmarking work of ref.\cite{Hennigar:2017apu} chose fixed $J$.) 

Fixing $a$   gives $dJ=adM$. So along the isobars ($dp=0$)  the total heat flow is  $\int TdS=\int dM - \int a\Omega dM$. It turns out that, perhaps surprisingly, for Kerr--AdS black holes,  $a\Omega dM$ is exactly integrable over $r_{+}$. For $D=4$, it is:
\bea \label{extraterm}
a\Omega dM=\frac{a^2\Big(1+\frac{r^2_{+}}{l^2}\Big)\Big(3r^4_{+}+(a^2+l^2)r^2_{+}-a^2 l^2\Big)}{2(a^2+r^2_{+})r^2_{+}l^2 \Xi ^2} dr_+\ .
\eea
Then integrating over $r_{+}$ keeping $l$ fixed results in:
\bea \label{newfn}
\Upsilon(r_{+},l,a)=\arctan \Big(\frac{r_{+}}{a}\Big) a +\frac{r_{+} a^2}{l^2 \Xi }\left[1+\frac{l^2}{2 r^2_{+} \Xi}\Big(1+\frac{2 r^2_{+}}{l^2}+\frac{r^4_{+}}{l^4}\Big)\right]\ .
\eea
 Appendix~\ref{Up} shows how this generalized to higher dimensions. This means that the efficiency formula  extends to:
\bea \label{ExtendedSimpleMass}
\eta=1-\frac{(M_{2}-M_{1})-(\Upsilon_{2}-\Upsilon_{1})}{(M_{3}-M_{4})-(\Upsilon_{3}-\Upsilon_{4})} \ ,
\eea
where $M_i$ and $\Upsilon_i$ are the value of those functions at each corner of the cycle. 

This extended result for a cell of type shown in figure~\ref{fig:cycles}(b) is what is used in the tessellation algorithm for computing the efficiency of the benchmark circle, with $\eta = 1-{Q_{C}}/{Q_{H}}$, where:
\bea \label{QH}
Q_{H}=\sum_{i{\rm th} \ {\rm hot} \ {\rm cell}} (M^{(i)}_{2}-M^{(i)}_{1})-(\Upsilon^{(i)}_{2}-\Upsilon^{(i)}_{1}) \ , \
\eea
\bea \label{QC}
 Q_{C}=\sum_{i{\rm th} \ {\rm cold} \ {\rm cell}} (M^{(i)}_{3}-M^{(i)}_{4})-(\Upsilon^{(i)}_{3}-\Upsilon^{(i)}_{4}) \ . \
\eea
Our benchmark circle will be centred at  $p_c=30$, $V_c=50$, with  radius $L=5$. For generating the moves along adiabats (as discussed in the previous section), we used equation (\ref{entropy}), setting $dS=0$ giving:
\bea \label{drdp}
dr_{+}=-\frac{4 \pi a^2(r^2_{+}+a^2)}{3r_{+}\Big(1-\frac{8 \pi p a^2}{3}\Big)}dp\ .
\eea
Then using equation (\ref{volume}) for $V$ and equation (\ref{drdp}) for $dr_{+}$, one gets:
\bea \label{dvdp}
dV=\frac{8 \pi ^2 a^4}{9r^3_{+}}(r^2_{+}+a^2)^2 \frac{\Big(1+\frac{8 \pi p r^2_{+}}{3}\Big)}{\Big(1-\frac{8 \pi p a^2}{3}\Big)^3}dp\ .
\eea
Before showing our results of implementing the algorithm, we pause to note an exact result that will be a useful guide.

\subsection{An Exact Result}
\label{sec:exact}

For static black holes, there is an ``ideal gas'' limit in which the equation of state in $D$ dimensions becomes form  $pV^{1/(D-1)}\sim T$. It is essentially a high temperature or large volume limit, and can be obtained by keeping the leading $r_+$ behaviour of the various expressions for $T$, $M$, {\it etc}. In this limit, the mass/enthalpy becomes $M=pV$ with an interesting consequence for the tessellation algorithm. Recall that the  input heat, $Q_H=\sum_k (M_2^{(k)}-M_1^{(k)})$ is summed along the tops of a stepwise discretization of the top of the benchmarking circle. But in this case it is $Q_H=\sum p(V)\Delta V$ along the  circle's top half. Therefore in the limit of small step size this becomes the area under the curve, a pleasing geometrical result which allows the formula~(\ref{eq:cjbound}) to be written down.

It is natural to wonder if a similar limit might apply here, allowing for another exact formula. The answer is yes. The large $r_+$ limit of the mass  is again $M=pV$, while the angular velocity  becomes $\Omega=a/l^2$. Therefore isobaric heat flow becomes:
\begin{equation}
\int TdS=\int\left(1-\frac{a^2}{l^2}\right)dM\longrightarrow\int\left(1-\frac{8\pi a^2}{3}p(V)\right)p(V)dV\ .
\end{equation}
So again the result is geometrical, with a weighting factor in the integral. So for a circle  of radius~$L$, centered at $p=p_c$ and $V=V_c$:
\begin{equation}
Q_H=\int_{V_c-L}^{V_c+L}\left(1-\frac{8\pi a^2}{3}p(V)\right)p(V)dV\ ,\quad {\rm where}\quad (p(V)-p_c)^2+(V-V_c)^2=L^2\ .
\end{equation}
The first (square root) term is the original result (the sum of the areas of the upper semi--circle and the rectangle), easily solved by the substitution $V=L\sin\theta$. 
Meanwhile the second term is an elementary polynomial. Recalling that $W=\pi L^2$,  the final ($D$--independent) result is:
\begin{equation}
\eta=\frac{2\pi}{\left[\pi+\frac{4p_c}{L}-\frac{16\pi}{3}a^2\left(\frac{4}{3}L+p_c\pi+\frac{2p_c^2}{L}\right)\right]}\ .
\label{eq:cjboundnew}
\end{equation}
As with the $a=0$ version, since it comes from an ideal gas limit, it probably expresses a limit on the efficiency achievable by a rotating black hole, at least in generic situations. Our results in the next section are consistent with this.

\subsection{Numerical Results}
\begin{figure}[h]
\begin{center}
{\centering
\includegraphics[width=3in]{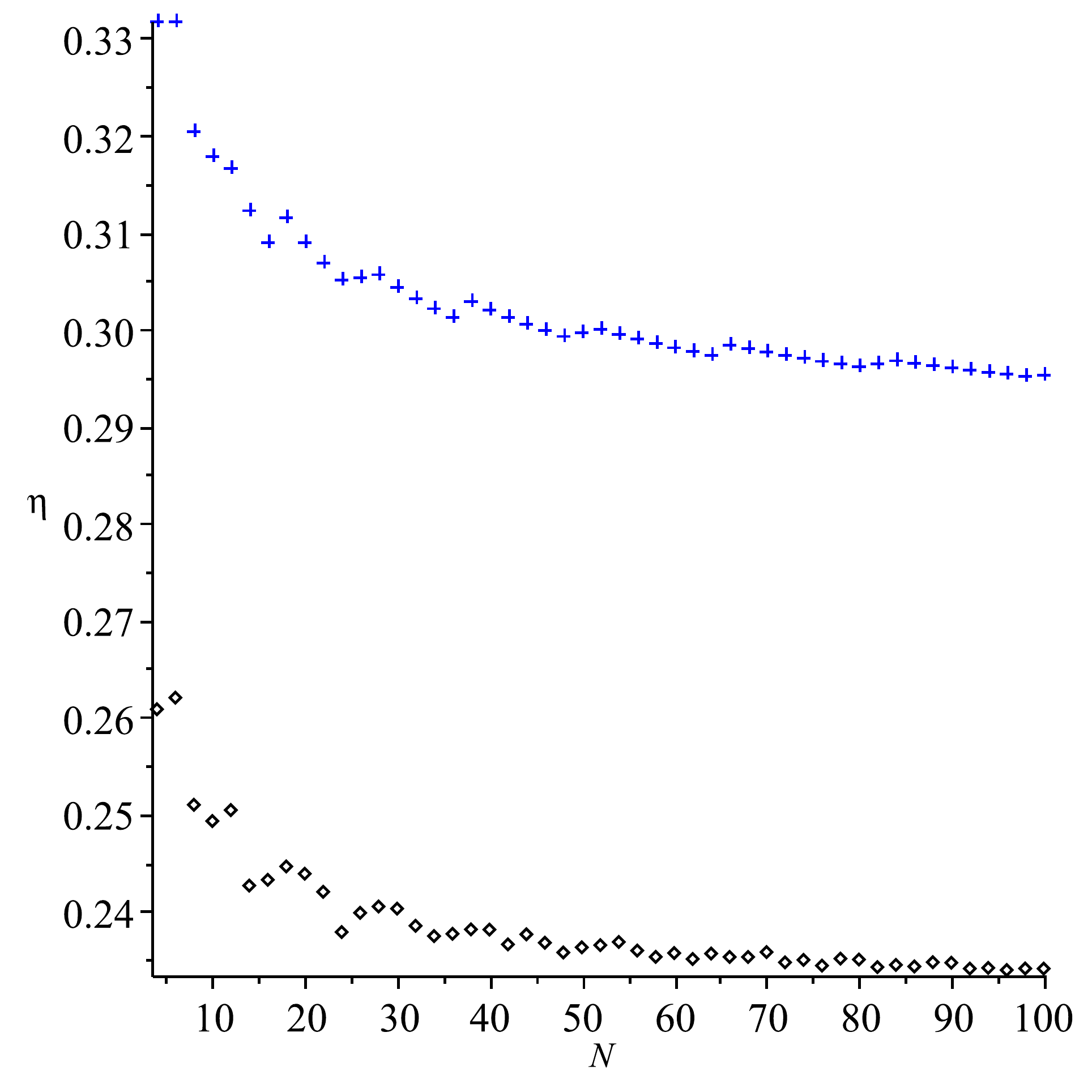} 
   \caption{\footnotesize  The efficiency of our benchmarking cycle as a function of grid size, $N$. Here Kerr--AdS black holes in $D=4$ are used as the working substance. Blue crosses represent the  Carnot efficiency $\eta_{\rm C}$, while black squares represent $\eta$. For $N=100$, $\eta_{\rm C}$ and $\eta$ begin to show convergence to approximately 0.295355525453494 and 0.234033334108944 respectively. We set $a=0.001$.}   \label{fig:loop}
}
   \end{center}
\end{figure}

Our benchmarking circle has origin at $(50,30)$ and radius $5$.  Figure~\ref{fig:loop} shows the results of the algorithm for computing  $\eta_{\rm C}$ and $\eta$. We worked with $a=0.001$ as a sample value of the rotation parameter\footnote{Recall that the metric is valid only for $a<l$ which sets an upper limit on $a$. With our chosen cycle, this means $a$ has to be less than $0.06$. We also have another restriction on $a$ coming from the extremality condition in equation~(\ref{eq:extremal}). For our choices of cycle $a_{\rm ext}$ is always greater than $0.06$. Notice that we also avoided the regions of the phase diagram where values of $a$ at which multi--valuedness associated with phase transitions would develop.}. 

\begin{figure}[h]
\begin{center}
{\centering
\includegraphics[width=3in]{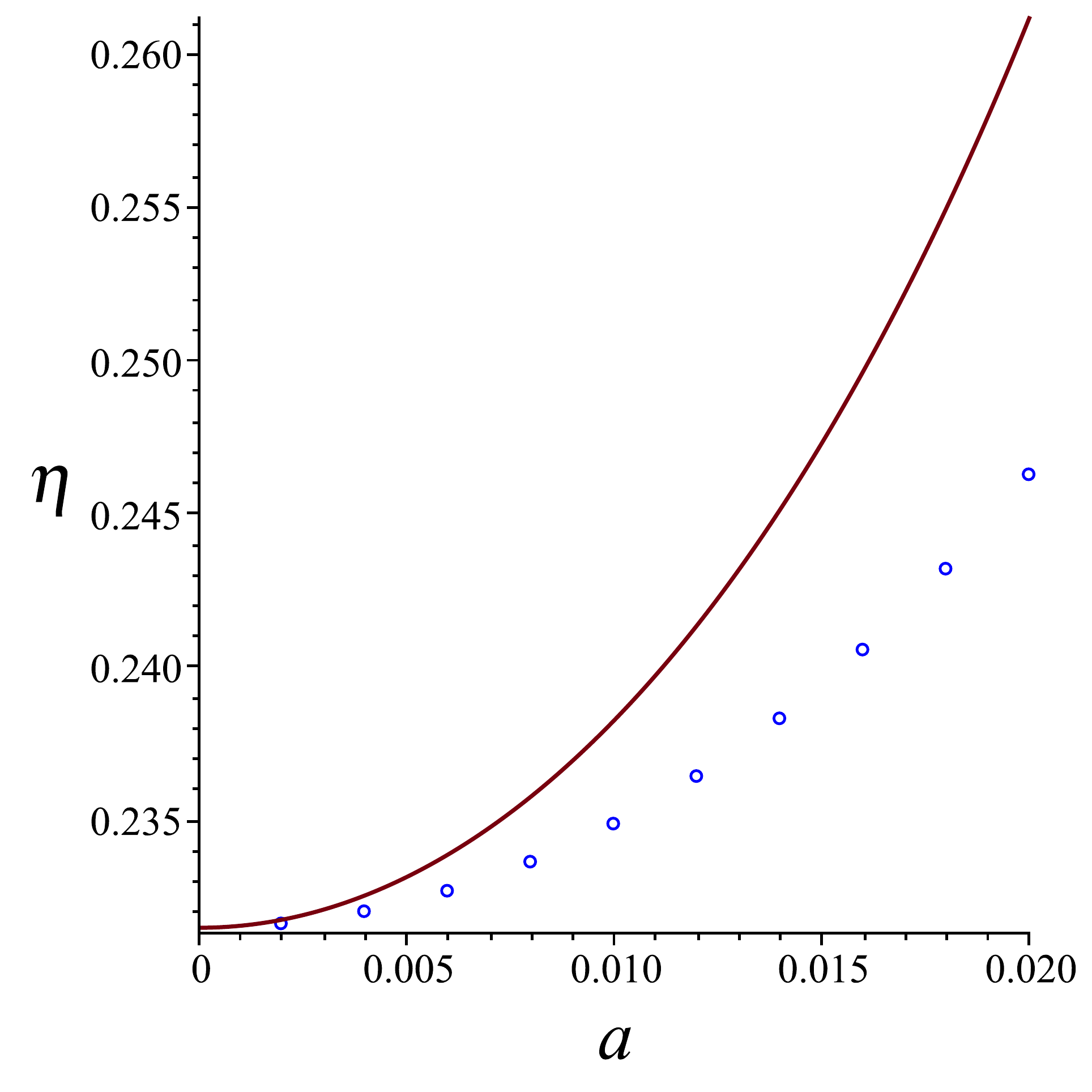} 
   \caption{\footnotesize  Results for the efficiency of the benchmarking circle for a range of rotation parameters. The results were obtained by direct integration of the heat to evaluate the efficiency. The blue dots are the numerical results, while the solid curve is the exact ``ideal gas'' result of equation~(\ref{eq:cjboundnew}).}   \label{fig:comparetoexact}
}
   \end{center}
\end{figure}

As $N$ grows, the tessellation scheme, while successful, becomes numerically labour intensive as compared to the static case of $a=0$. That difficult grows with larger $a$. Being able to do large enough $N$ for a range of values of $a$, to ensure reliable convergence,  eventually became beyond the capabilities of our computers. Instead, for a study of the $a$--dependence of the system, we relied on a direct numerical integration of the heat around the cycle in order to evaluate the efficiency at different  $a$. The results are shown in figure~\ref{fig:comparetoexact} as a series of points indicated by circles. The solid (red) line is the exact result (\ref{eq:cjboundnew}) coming from the ideal gas case and again it seems to be consistent with our expectation that it is an upper bound.

\begin{figure}[h]
\begin{center}
{\centering
\includegraphics[width=5.5in]{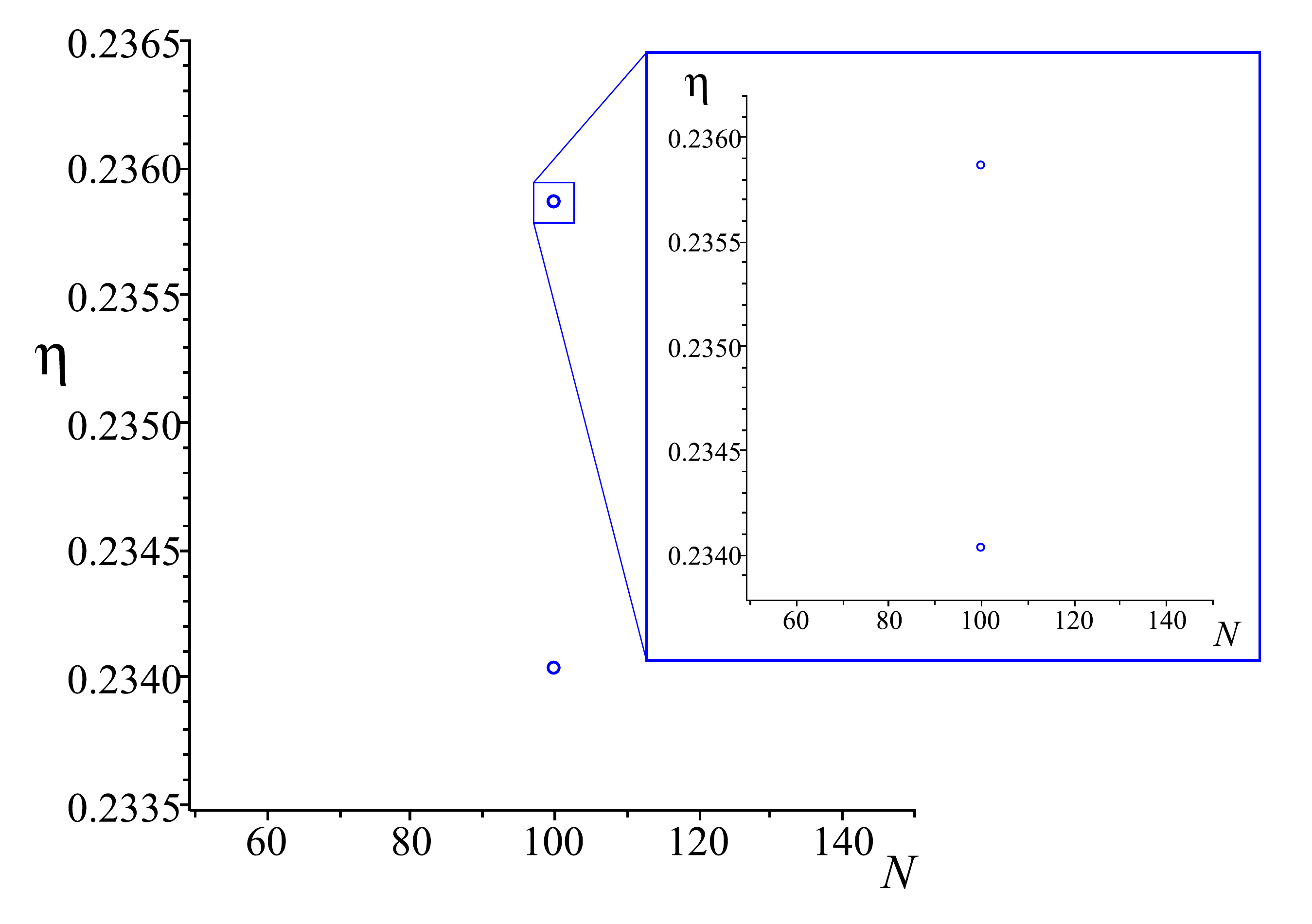} 
   \caption{\footnotesize   The efficiencies of Kerr--AdS (lowest), Born--Infeld (middle) and Einstein--Hilbert--Maxwell (highest) black hole heat engines for $N=100$ with circle origin at $(50,30)$ and radius $L=5$. (Since the static cases lie close to each other in the plot, an inset is included to resolve them.)}   \label{fig:compare}
}
   \end{center}
\end{figure}

We next compare the efficiencies of Kerr--AdS black hole heat engines with that of the static Einstein-Hilbert--Maxwell and Born--Infeld cases.  In the interests of getting better numerical control, we used a  different  choice of benchmarking  circle (origin at $(V_c=50,p_c=30)$ and radius $5$) than we did in ref. \cite{Chakraborty:2016ssb} (we had an origin at $(V_c=20,p_c=110)$ and radius 10), and there we worked in $D=5$, so we must redo the static results.  We list the actions (and mass formulae needed)  in Appendix~\ref{BHs}. In $D=4$, the Gauss--Bonnet term is purely topological and so we could  not include that  case. We performed our computation for $N=100$ and the result is shown in Figure~\ref{fig:compare}.  We chose $\beta=0.1$ for the Born--Infeld parameter, and display the case of electric charge $q=0.1$ in the figure. (Other values of $q$ are discussed in the next paragraph.)  Notably,  inclusion of rotation results in lower efficiency than the static cases. In $D=4$,  Born--Infeld case has lower efficiency than the Einstein--Hilbert--Maxwell case (as opposed to the $D=5$ case), and Kerr--AdS has the lowest efficiency ($a$ was set to $0.001$), and of course all of them have efficiencies lower than their corresponding Carnot efficiencies. Note that in \cite{Hennigar:2017apu} it was   found that in $D=4$, Born--Infeld black hole heat engines have higher efficiencies than the Einstein--Hilbert--Maxwell heat engines (the reverse of what we found) for values of $q$ below some threshold value. This does not necessarily contradict our results here since the systems are in  different thermodynamic ensembles. Nonetheless, it would be interesting to find an explanation for the origin of this reversal.  

 Of course, for small enough values of the charge parameter $q$, the difference between the  Einstein--Hilbert--Maxwell and Born--Infeld heat engines  efficiencies  becomes so small as to be lost in numerical noise. We  then explored the behaviour of the efficiencies as we increase the value of $q$, until our numerical accuracy became too unreliable. We found that the  Einstein--Hilbert--Maxwell engine efficiency increases with $q$ at a faster rate than that of Born--Infeld. Ultimately, the  Born--Infeld engine efficiency reached a maximum  (at $q\approx0.5$ in the units we are using) and then started decreasing. There is no such peak for Einstein--Hilbert--Maxwell. Presumably this means that  the effects of non--linearity for Born--Infeld eventually begin to take their toll on the efficiency.
 
\section{Conclusion}
\label{Conc}

We have extended our  benchmarking scheme \cite{Chakraborty:2016ssb} to compute the efficiency of non--static AdS black hole heat engines, allowing us to study the effects of rotation on black hole heat engine efficiency. Specifically, we did it for four dimensional Kerr--AdS black holes. We used our modified algorithm and exact efficiency formula (\ref{ExtendedSimpleMass}) to break the circle into a regular lattice of cycles and compute the efficiency numerically. In the special case of an ``ideal gas'' behaviour we were able to find a generalization of the exact efficiency formula we'd previously derived in ref. \cite{Chakraborty:2016ssb}, again proposing that it supplies a stronger bound than Carnot, now for the rotating black holes (in a fixed angular momentum ensemble).  

We also computed efficiencies of Einstein--Hilbert--Maxwell and Born--Infeld charged black hole heat engines in $D=4$ and compared those with the rotating case. We  showed in ref. \cite{Chakraborty:2016ssb} that Born--Infeld yields higher efficiencies in $D=5$ in this benchmarking scheme (note that we double--checked that this result persisted with the circle that we used in this paper).  Here instead,  we found that in $D=4$ we have the reverse: Now Born--Infeld has lower efficiency than Einstein--Hilbert--Maxwell case. Furthermore,  the  inclusion of rotation results in efficiency lower than that of both the static cases discussed here. It would be interesting to know if there was a simple explanation for the dimension dependence of the behaviour of the efficiencies. Moreover, there are several other kinds of black hole that can be compared using our benchmarking scheme. These explorations will be left for future work.

\section*{Acknowledgments}

We would like to thank   the  US Department of Energy for support under grant DE--SC0011687. This work was partially supported by a grant from the Simons Foundation (\#395705 to Clifford Johnson). CVJ  would also like to thank Amelia for her patience and support.


\appendix

\section{Black Holes in $D=4$}
\label{BHs}

Here we write down the mass and temperature of Einstein--Hilbert--Maxwell and Born--Infeld black holes, which we were previously using in our benchmarking study  in $D=5$  and with a different choice of benchmarking circle for static black holes  \cite{Chakraborty:2016ssb}. 

\subsection{Einstein--Hilbert--Maxwell}
\label{EHM}

The bulk action for the Einstein--Hilbert--Maxwell system in $D=4$  is\footnote{We're using the conventions of ref. \cite{Chamblin:1999tk}.}:
\bea \label{EHMaction}
I=\frac{1}{16\pi}\int d^4 x\sqrt{-g}\Big(R-2\Lambda -F^2\Big) \ .
\eea 

We can now write the mass and the temperature of the Einstein--Hilbert--Maxwell ({\it i.e., } Reissner--Nordstrom--like) black hole solution, parametrized by a charge $q$ (which we will  choose  as~$q=~0.1$):

\bea
M=\frac{1}{2}\Big(r_{+}+\frac{q^2}{r_{+}}+\frac{8 \pi p}{3}r^3_{+}\Big)\ ,\quad {\rm and} \quad
T=\frac{1}{4\pi}\Big(8 \pi p r_{+}+\frac{1}{r_{+}}-\frac{q^2}{r^3_{+}}\Big)\ , 
\eea
 and we can write them entirely in terms of $p$ and $V$, using  $r_{+}=(3V/4\pi)^{1/3}$. 




\subsection{Born--Infeld}
\label{BI}

The so--called\footnote{See {\it e.g.} the remarks in ref.\cite{Johnson:2015fva} about the terminology} Born--Infeld action \cite{Born:1933,Born:1934ji,Born:1934gh} is a non-linear generalization of the Maxwell action, controlled by the parameter $\beta$ :
\bea \label{BIsector}
{ \cal L}(F)=4 \beta^2 \Big(1-\sqrt{1+\frac{F^{\mu \nu}F_{\mu \nu}}{2 \beta^2}}\Big)
\eea
If we take the limit $\beta \rightarrow \infty$ in (\ref{BIsector}) we recover the Maxwell action. The Einstein--Hilbert--Born--Infeld bulk action in $D=4$ is obtained by replacing the Maxwell sector in equation (\ref{EHMaction}) with this action.
The exact results for the Born--Infeld black hole's mass and temperature are known\footnote{See refs. \cite{Fernando:2003tz,Cai:2004eh,Dey:2004yt} for further details.}, but for our purposes, it is enough to expand them in $1/\beta$, keeping only  leading non--trivial terms. For the mass:
\bea
M=\frac{1}{2}\Big(r_{+}+\frac{8 \pi p}{3}r^3_{+}+\frac{q^2}{r_{+}}(1-\frac{2 q^2}{15 \beta^2 r^4_{+}})\Big)+\mathcal{O}\Big(\frac{1}{\beta^4}\Big)
\eea
and the temperature:
\bea
T=\frac{1}{4 \pi}\Big(\frac{1}{r_{+}}+8 \pi p r_{+} - \frac{q^2}{r^3_{+}}(1-\frac{q^2}{4 \beta^2 r^4_{+}})\Big)+\mathcal{O}\Big(\frac{1}{\beta^4}\Big)
\eea
 We  worked with $q=0.1$ and $\beta=0.1$ in our benchmarking scheme.

\section{$\Upsilon$-function in higher dimensions}
\label{Up}

The $\Upsilon$-function that we have written down explicitly in section \ref{sec:kerr} was derived for $D=4$. In this appendix we will show that it can also be done for higher dimensional singly spinning Kerr--AdS black holes. A singly spinning Kerr--AdS black hole in general $D$-dimensions can be described by one non-zero rotation parameter $a$ and the metric takes the form:
\bea
ds^2&=&-\frac{\Delta_{r}}{\rho ^2}\Big[dt-\frac{a \sin^2 \theta}{\Xi} \ d\phi \Big]^2+\frac{\rho ^2}{\Delta_{r}} \ dr^2+\frac{\rho ^2}{\Delta_{\theta}} \ d\theta ^2\nonumber \\
&&\hskip 4cm +\frac{\Delta_{\theta} \ \sin^{2}\theta}{\rho ^2} \Big[a dt-\frac{r^2+a^2}{\Xi} \ d\phi \Big]^2+r^2 \cos^2 \theta d\Omega^2_{D-4} \ ,
\eea
where,
\bea
\rho^2 =r^2+a^2 \cos^2\theta \ , \ \Xi =1-\frac{a^2}{l^2} \ ,
\eea
\bea \label{radialD}
\Delta_{r}=(r^2+a^2)\Big(1+\frac{r^2}{l^2}\Big)-2mr^{5-D} \ , \ \Delta_{\theta}=1-\frac{a^2}{l^2} \cos^2\theta \ .
\eea
Other useful thermodynamic quantities are:
\bea \label{Ang}
M=\frac{m\omega_{D-2}}{4 \pi \Xi ^2}\Big(1+\frac{(D-4)\Xi}{2}\Big) \ , \ 
J=\frac{m a \omega_{D-2}}{4 \pi \Xi ^2} \ , \ 
\eea
\bea \label{AngVel}
V=\frac{r_{+}A}{D-1}\Big[1+\frac{a^2 (1+\frac{r^2_{+}}{l^2})}{(D-2)r^2_{+}\Xi}\Big] \ , \ 
\Omega = \frac{a\Big(1+\frac{r^2_{+}}{l^2}\Big)}{r^2_{+}+a^2} \ .
\eea
where,
\bea
A=\frac{\omega_{D-2}(r^2_{+}+a^2)r^{D-4}_{+}}{\Xi} \ , \quad {\rm and} \quad S=\frac{A}{4} \ , 
\eea
and $\omega_{D-2}= 2 \pi ^{(\frac{D-1}{2})}/\Gamma(\frac{D-1}{2})$ is the usual volume of the unit $(D-2)$-sphere. 
As before, $p=-\frac{\Lambda}{8 \pi}$, where $\Lambda$ is the cosmological constant and related to $l$ by $\Lambda = -\frac{(D-1)(D-2)}{2 l^2}$.
The statement that $r_{+}$ is the largest root of $\Delta_r$ gives us mass $M$ in term of the horizon radius, $r_{+}$:
\bea \label{MassD}
M=\frac{\omega_{D-2}}{4 \pi \Xi ^2}\Big(1+\frac{(D-4)\Xi}{2}\Big)\Big[\frac{(r^2_{+}+a^2)(1+\frac{r^2_{+}}{l^2})}{2r^{5-D}_{+}}\Big]
\eea
Now we can easily compute $dJ$ from (\ref{Ang}) using equation (\ref{MassD}) (while keeping $p$, i.e., $l$ constant). Next, we multiply $dJ$ by $\Omega$ using (\ref{AngVel}). This quantity, $\Omega dJ$, is integrable exactly along the isobars in any $D$. For general $D$, $\Omega dJ$ takes the following form:
\bea \label{GenD}
\frac{a^2 \omega_{D-2}(1+\frac{r^2_{+}}{l^2})\Big[(D-5)(a^2+r^2_{+})(1+\frac{r^2_{+}}{l^2})r^{D-6}_{+} + 2(1+2\frac{r^2_{+}}{l^2}+\frac{a^2}{l^2})r^{D-4}_{+}\Big]}{8 \pi (a^2+r^2_{+})\Xi ^2} \, dr_+ \ .
\eea
The result upon integrating (\ref{GenD}) is what we called $\Upsilon$ (for $D=4$) in section \ref{sec:kerr}. We present two such results here. In $D=5$ , $\Upsilon$ takes the form:
\bea
\Upsilon_5 =\frac{a^2 \omega_3}{8 \pi}\Big[ \ln(a^2+r^2_{+})+\frac{r^2_{+}}{l^2 \Xi}\Big(1+\frac{2}{\Xi}+\frac{r^2_{+}}{l^2 \Xi}\Big)\Big]
\eea
and in $D=6$ , the result is:
\bea
\Upsilon_6 =\frac{a^2 \omega_4}{4 \pi}\Big[\frac{r_{+}}{l^2 \Xi}\Big(r^2_{+}(\frac{1}{3}+\frac{1}{\Xi})+(l^2-a^2)+\frac{(l^4+r^4_{+})}{2 l^2 \Xi}\Big)-\arctan(\frac{r_{+}}{a})a\Big]
\eea
One can derive exact results for higher dimensions too using (\ref{GenD}).

%
\bibliographystyle{utphys}
\bibliography{Johnson_rotating}




\end{document}